\documentclass[twoside,english,10pt]{JAC2001}

\usepackage{amssymb}
\usepackage{graphicx}
\usepackage{babel}
\usepackage{multicol}
\usepackage[centerlast]{caption2}

\begin{document}

\title{Autopsy on an RF-Processed X-band Travelling Wave Structure}
\author{F. Le Pimpec, S. Harvey, R.E. Kirby, F. Marcelja
\\SLAC, 2575 Sand Hill Road, Menlo Park CA 94025 , USA
}
\maketitle

\begin{abstract}

In an effort to locate the cause(s) of high electric-field
breakdown in x-band accelerating structures, we have
cleanly-autopsied (no debris added by post-operation structure
disassembly) an RF-processed structure. Macroscopic localization
provided operationally by RF reflected wave analysis and acoustic
sensor pickup was used to connect breakdowns to autopsied crater
damage areas. Surprisingly, the microscopic analyses showed
breakdown craters in areas of low electric field. High currents
induced by the magnetic field on sharp corners of the input
coupler appears responsible for the extreme breakdown damage
observed.

\end{abstract}


\section{Introduction}

\begin{figure}[htbp]
\centering
\vspace{-0.2cm}
\includegraphics[clip=,width=7.5cm,totalheight=5cm]{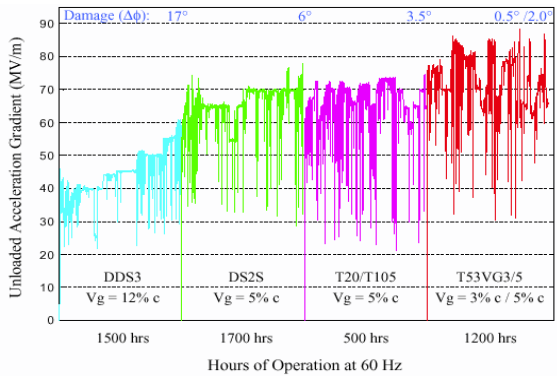}
\vspace{-0.4cm}
\caption{History operation at NLCTA of some 11.424 GHz TW structures}
\vspace{-0.2cm}
\label{historyOperation}
\end{figure}

\begin{figure}[htbp]
\vspace{-0.2cm}
\begin{minipage}[t]{.5\linewidth}
\centering
\includegraphics[clip=,width=4cm,totalheight=4cm]{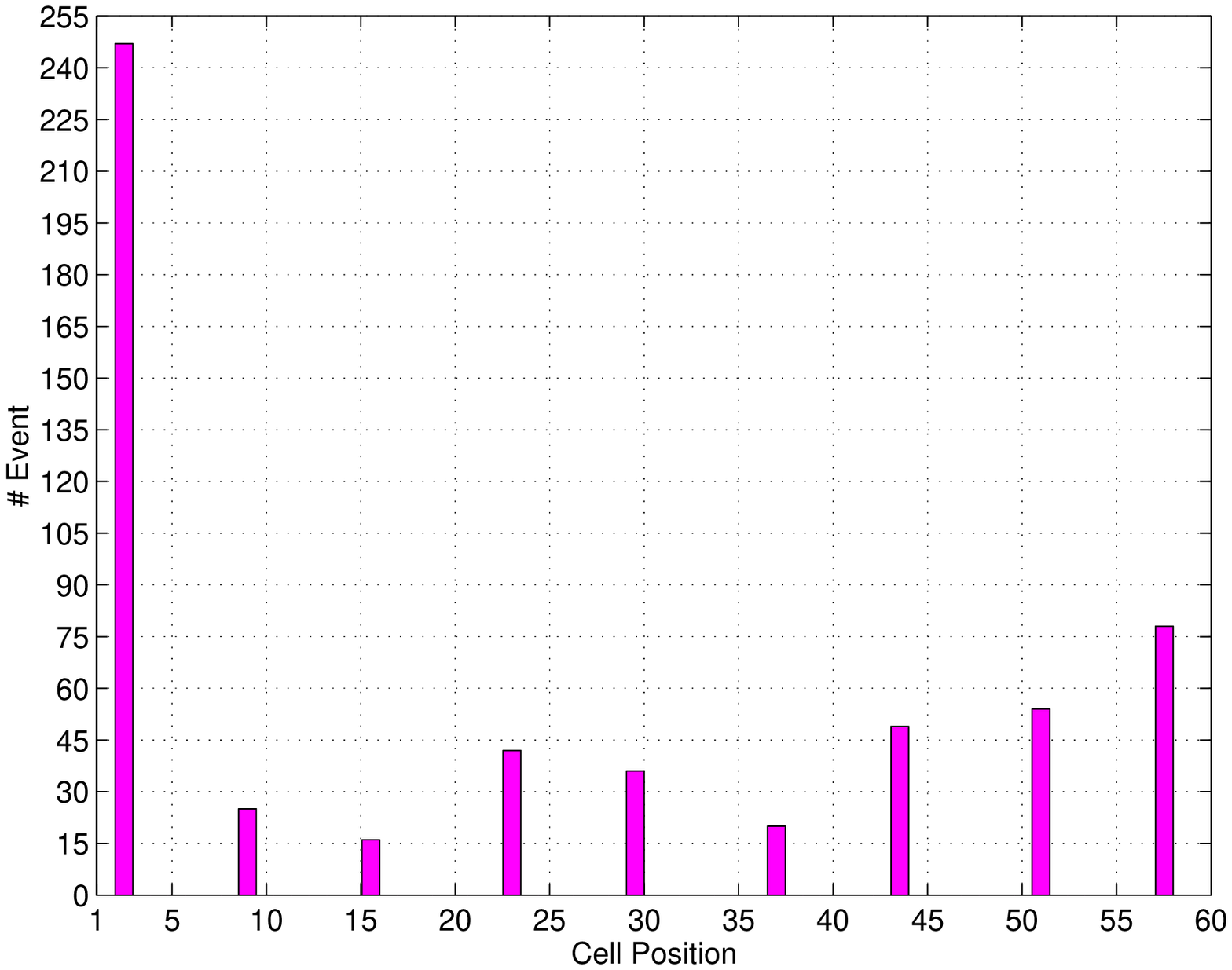}
\end{minipage}%
\begin{minipage}[t]{.5\linewidth}
\centering
\includegraphics[clip=,width=4cm,totalheight=4cm]{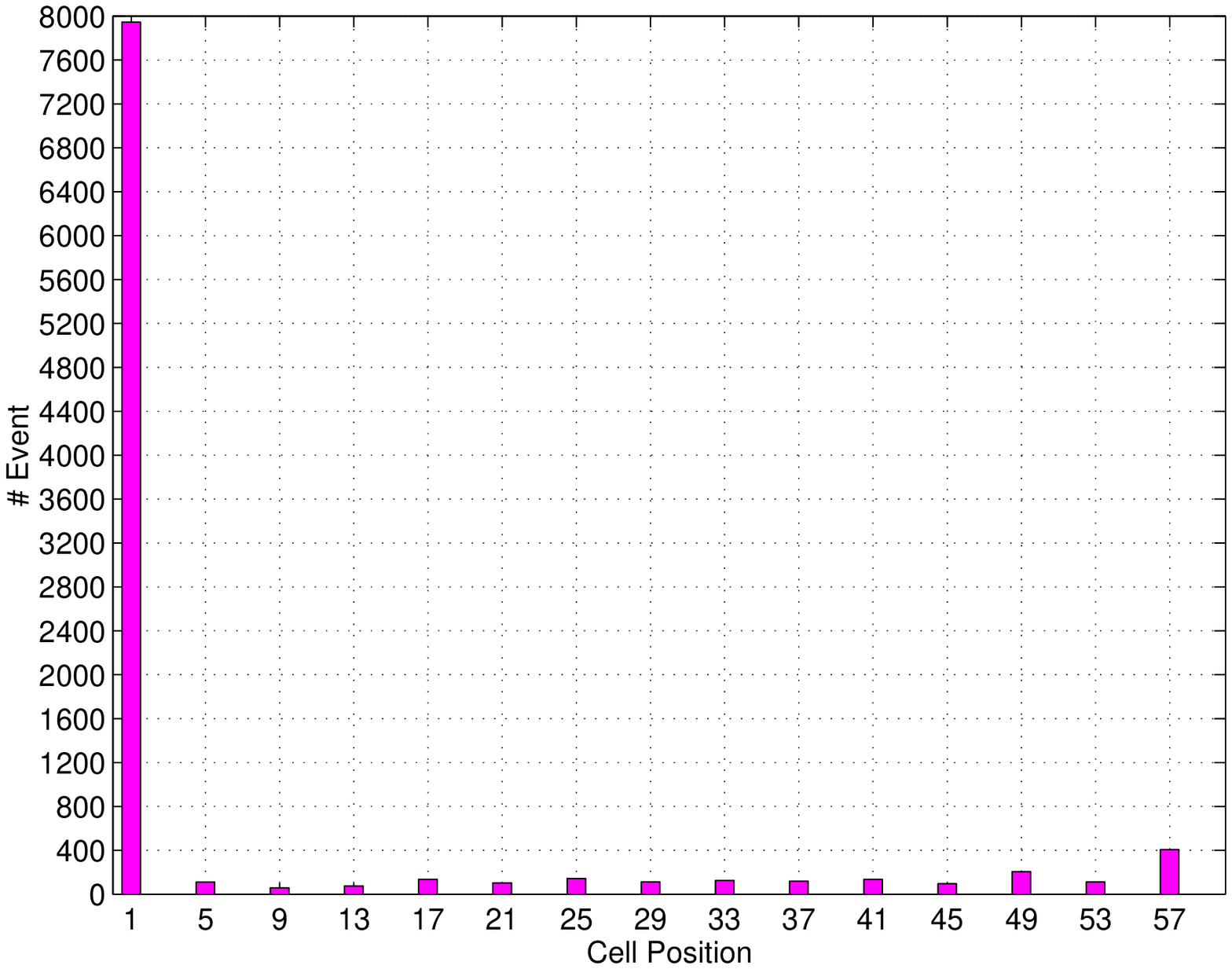}
\end{minipage}
\setcaptionwidth{7.5cm} \caption{Example of damage localization in
T53Vg3R (left) - T53Vg3RA (right) during operation}
\vspace{-0.2cm} \label{T53Vg3}
\end{figure}

The Next Linear Collider (NLC) accelerating structures, running at
11.424 GHz, are expected to hold a steady operational surface
gradient of 73 MV/m, without breakdown arcing. A development
program for the required structure has been in progress at SLAC
for several years. Possible RF structure candidates, Travelling
Wave (TW) and Standing Wave (SW) are tested in a specially
dedicated linac at SLAC, the Next Linear Collider Test Accelerator
(NLCTA). So far, mixed results have been obtained during high
electric field-processing of structures, cf
Fig.~\ref{historyOperation}. Part of the success in reaching the
required gradient has been the great improvement in cleaning and
handling the structure prior to RF processing. Results of the
breakdown pattern analysis via RF or acoustic sensors
Fig.\ref{T53Vg3}, 
have led to the necessity to cut open the structure. In this paper
we will present some of the results of the analysis of the autopsy
of the 53~cm long T53Vg3R TW structure.

\section{Results of the Analysis and Discussion}

During RF breakdown few kA/A of electrons/ions are emitted inside
a cell. The kinetic energy of electrons can be up to 100~keV and
ions up to few keV  \cite{dolgashev:PAC01} \cite {dolga:01}. Among
the interactions of those particles with the surface, few of them
might directly cause damage to the surface such as surface heating
and sputtering.

\begin{figure}[htbp]
\centering
\vspace{-0.2cm}
\includegraphics[clip=,width=6.5cm,totalheight=5cm]{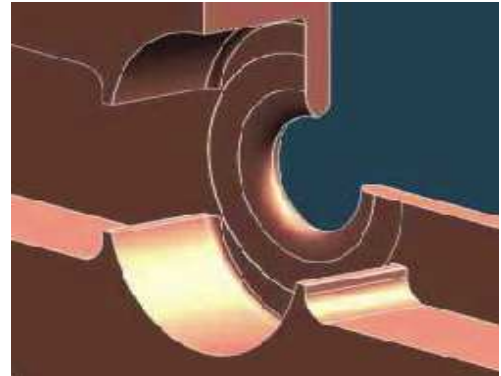}
\vspace{-0.4cm}
\caption{Input coupler design. The center is the cavity side with the first iris}
\vspace{-0.2cm}
\label{DesignIC}
\end{figure}

The breakdown pattern localization Fig.\ref{T53Vg3}
reveals that most of the breakdown are occurring in the input
coupler (IC) of the TW. In Fig.\ref{DesignIC} we can see that the
input coupler is divided into two parts the waveguide section,
left side of the picture and the cell side in the center. The
feature separating the waveguide side and the cell side is
denominated "horns". The gap between the top and bottom horn is
$\sim$~9mm.

\begin{figure}[htbp]
\centering
\vspace{-0.2cm}
\includegraphics[clip=,width=7.5cm,totalheight=5cm]{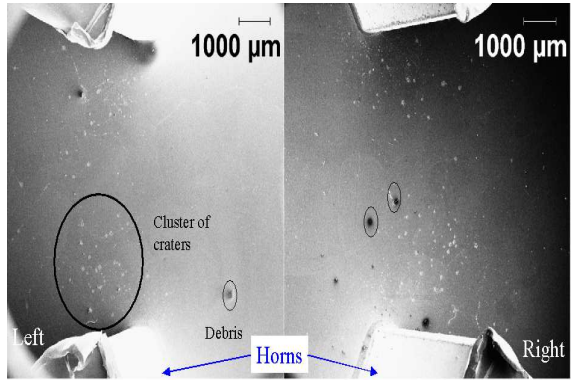}
\vspace{-0.4cm}
\caption{Pitting due to RF breakdown on the floor of the Input coupler.
The iris is in between the horns.}
\vspace{-0.2cm}
\label{ICfloor}
\end{figure}

\begin{figure}[htbp]
\centering
\vspace{-0.2cm}
\includegraphics[clip=,width=7.5cm,totalheight=5cm]{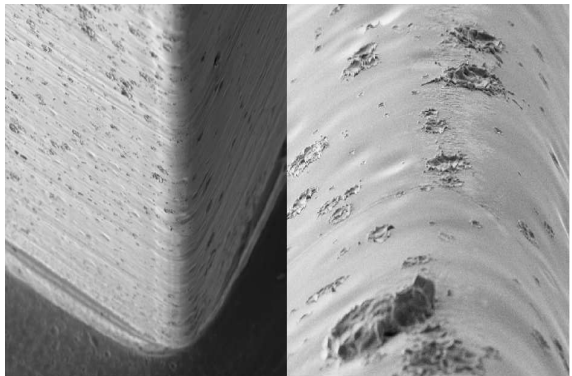}
\vspace{-0.4cm}
\caption{Typical damage on the waveguide side of a horn of the IC}
\vspace{-0.2cm}
\label{ICWGdamage}
\end{figure}

\begin{figure}[htbp]
\centering
\vspace{-0.2cm}
\includegraphics[clip=,width=7.5cm,totalheight=5cm]{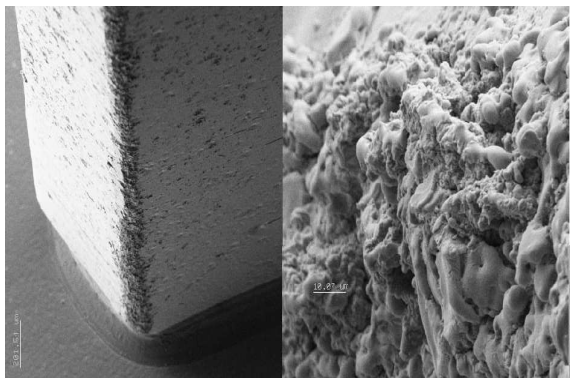}
\vspace{-0.4cm}
\caption{Typical damage on the Cell side of a horn of the IC}
\vspace{-0.2cm}
\label{ICcelldamage}
\end{figure}

Fig.\ref{ICfloor}, \ref{ICWGdamage} and \ref{ICcelldamage} reveal
a very interesting pattern of damage. The craters on the floor of
the IC, Fig.\ref{ICfloor}, are mainly located along the imaginary
circles connecting the two faces of the horns. Closing up to the
horns and focusing on the details, one can observe droplets of
copper \cite{lepimpec:ISG8}. Craters appears on all the sides of
the horns, with a density higher in the face side than the
waveguide or cavity side. Looking more closely of the edges of the
four horns has revealed that the edge on the waveguide side,
Fig.\ref{ICWGdamage}, shows little damage. In the contrary the
edge on the cell side shows a chaos of features, Fig.
\ref{ICcelldamage}, with sharp and melted objects. The dimension
bar in the right picture of Fig.\ref{ICcelldamage} is $\sim 10\mu m$.

\begin{figure}[htbp]
\centering
\vspace{-0.2cm}
\includegraphics[clip=,width=8cm,totalheight=5cm]{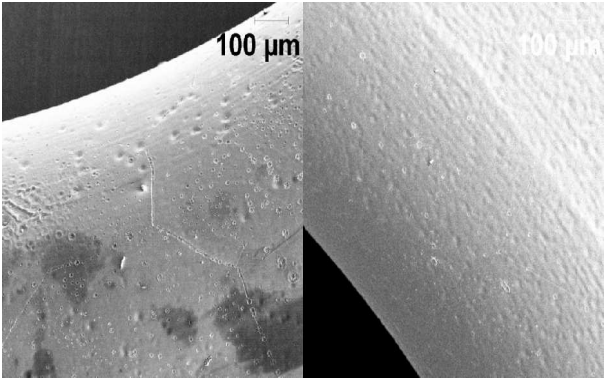}
\vspace{-0.4cm}
\caption{Activity on the 1$^{st}$ iris, Upstream side left picture and
downstream side right picture}
\vspace{-0.2cm}
\label{1stcell}
\end{figure}

\smallskip
The observation of the damages on the horns is coherent with the
acoustic sensors analysis. Also pictures taken during breakdown
show clearly after averaging over 100 of breakdowns 4 spots in the
approximative location of the horns \cite{mross:LC02}. At first
Ion bombardement and sputtering might be invoked to explain the
damages. However, the horns lies in a very low electric field
area, and the vacuum is of the order of 3.10$^{-9}$~Torr. It is
then not possible to explain how such a low gas density and
kinetic energy might start a breakdown leading to the so seen
damages. It is also not possible to explain the very localization
on the edge of the cavity side of the horns
Fig.\ref{ICcelldamage}. During processing, the RF pulse went from
50~ns to 400~ns carrying up to 100~MW of power. Part of this
energy is dumped in the Cu in form of heat or directly or by the
induced current due to the high magnetic field 0.7~MA/m at 400~ns
pulse length \cite{dolgashev:isg8}. Along the edge (76~$\mu m$
radius) this RF pulse heating leads to an increase of temperature
up to 130~$^{\circ}$C. It is clear that this temperature do not
cause the observed damage. However, it might induce thermal
fatigue in the Cu, inducing formations of cracks, as observed in
\cite{pritzkau}, and leading in roughening of the edge. As the
surface becomes rougher, the local resistivity of the Cu increases
leading to a higher temperature maybe up to the melting point
(1083~$^{\circ}$C). At this temperature, the vapor pressure of the
Cu is $\sim 3.10^{-4}$~Torr \cite{berman:92}. This gas can now be
ionized by the electrons coming form the surface and back bombard
the horns. As a result, raising up the temperature and sputtering
the surface. We might, in addition, explain the droplet of Cu seen
at the surface of the floor. The patterns of craters on the floor,
Fig.\ref{ICfloor}, seems to follow the high magnetic field line.
As we move toward the center the density of craters diminish and
increase again as the electric field start to be consequent
Fig~\ref{1stcell}.

\smallskip
SEM (secondary electron microscopy) Fig.~\ref{1stcell} shows the
damage on the upstream side and the downstream side of the iris,
respectif to the displacement of the NLCTA electron beam. On both
sides etch pits can be seen as well as craters. However, the
density of craters is much higher in the upstream side. Many
craters are also located in the grain boundary. A particle search
in and out of the craters shows MnS located in and out of them.
Most of this inclusions are located in the grain boundary. The
analysis of the contents of the craters shows mainly pure Cu. The
black spots on the upstream side is carbon. An Auger analysis of
the surface, not including those spots, shows heavy contamination
by carbon of this iris \cite{lepimpec:ISG8}. Carbon contamination
due to electron bombardement is a well known problem,
\cite{sciffmann:93}. There is no such contamination in iris 28
further down the structure \cite{lepimpec:ISG8}.

\begin{figure}[htbp]
\centering
\vspace{-0.2cm}
\includegraphics[clip=,width=8cm,totalheight=5cm]{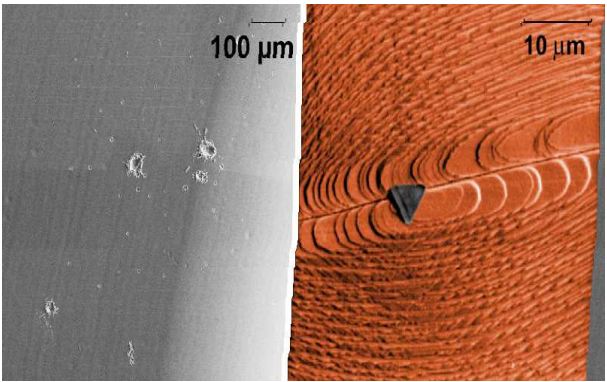}
\vspace{-0.4cm}
\caption{Activity on the 28$^{th}$ iris, Upstream side.
CrS inclusion at a twin boundary, right picture}
\vspace{-0.2cm}
\label{28cell}
\end{figure}

\smallskip
As we move toward the output coupler (OC), the activity on the irises
decreases, endoscope (boroscope) monitoring. Fig.\ref{28cell} shows
the sparse activity on the 28th iris of the structure. As we
looked for particles/inclusions on the surface and at the grain
boundary, we did not find MnS. However, some other sulfur compound
were found \cite{lepimpec:ISG8}, as the atypical one shown in
Fig.\ref{28cell}.

\smallskip
The cells used to make the structure are bonded and then brazed in
a stainless steel (SS) H$_2$ furnace at $\sim 1020~^{\circ}$C. The
Sulfur (S) composition is 0.03 weight \% in stainless steel. Other
material like Mn, Cr, Ni are also used in the making of SS. The
solubility of sulfur in copper is very high at this temperature
and significant at a few hundred~$^{\circ}$C. Hence, a significant
amount of S can get into the copper. After its fabrication, the
structure is vacuum baked in a SS pot at 650$^{\circ}$C for few
days, in order to desorb the H$_2$. Metals might be present in the
atmosphere of the H$_2$ furnaces and bakeout pot at this given
temperature, with vapor pressures of 10$^{-5}$ Torr for Mn,
10$^{-11}$~Torr for Cr, 10$^{-1}$~Torr for Ca and below
10$^{-11}$~Torr for Ni \cite{berman:92}. At 300$^{\circ}$C and
above the S concentrates at the grain boundaries of the Cu. The S
also start migrating to the surface at low temperature
75$^{\circ}$C \cite{Singh:80}. The Model for sulfur compound
formation is that the sulfur getters the metal vapor out of the
furnace atmosphere to form crystals/inclusions at/near the grain
boundaries. A particle search on the IC reveals that $\sim$80\% of
the particles are Mn-S. On the very next iris the proportion falls
to $\sim$50\%. The search in the 28$^{th}$ cell shows no Mn-S but
S-compounds are present at $\sim$50\%. Finally the search on the
last cell in front of the OC shows Mn-S and other S-compounds.
Finally in all the cases, other particles are compounds based on
C, Al and Si not excluding the presence of Mn or S or other
elements. This model might explain the sulfur compound formation,
but the understanding of the distribution along the structure of
those elements is under investigation.

\section{Conclusion}

\begin{figure}[htbp]
\vspace{-0.2cm}
\begin{minipage}[t]{.5\linewidth}
\centering
\includegraphics[clip=,width=4cm,totalheight=4cm]{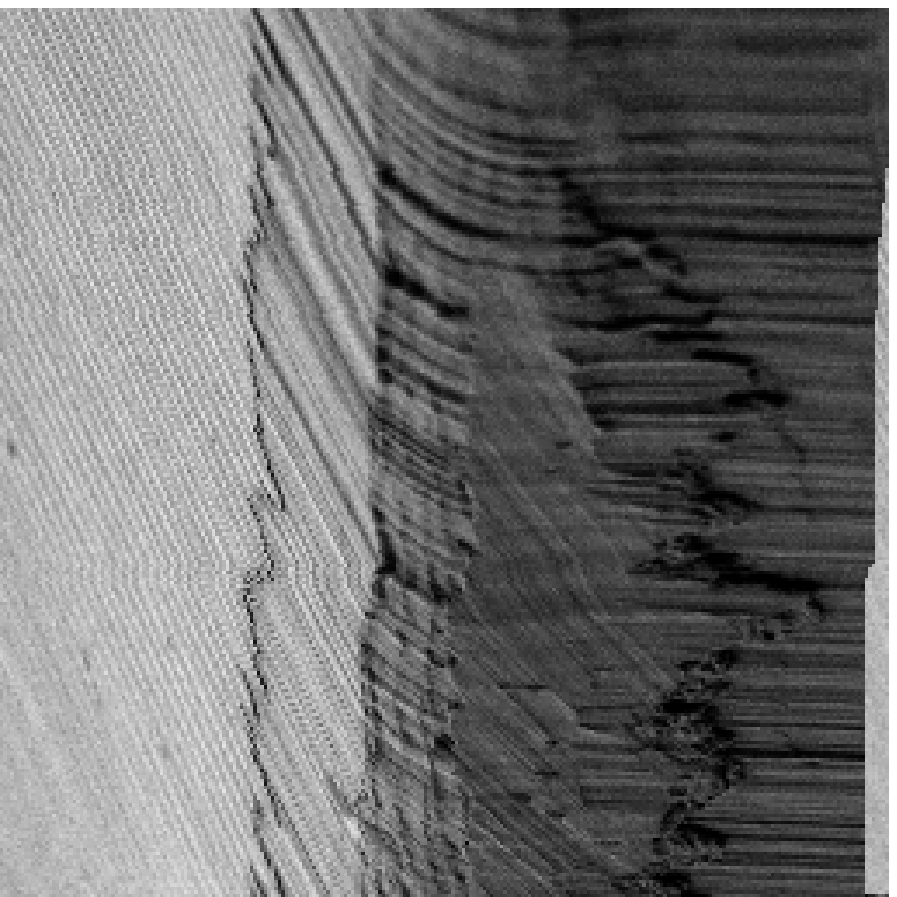}
\end{minipage}%
\begin{minipage}[t]{.5\linewidth}
\centering
\includegraphics[clip=,width=4cm,totalheight=4cm]{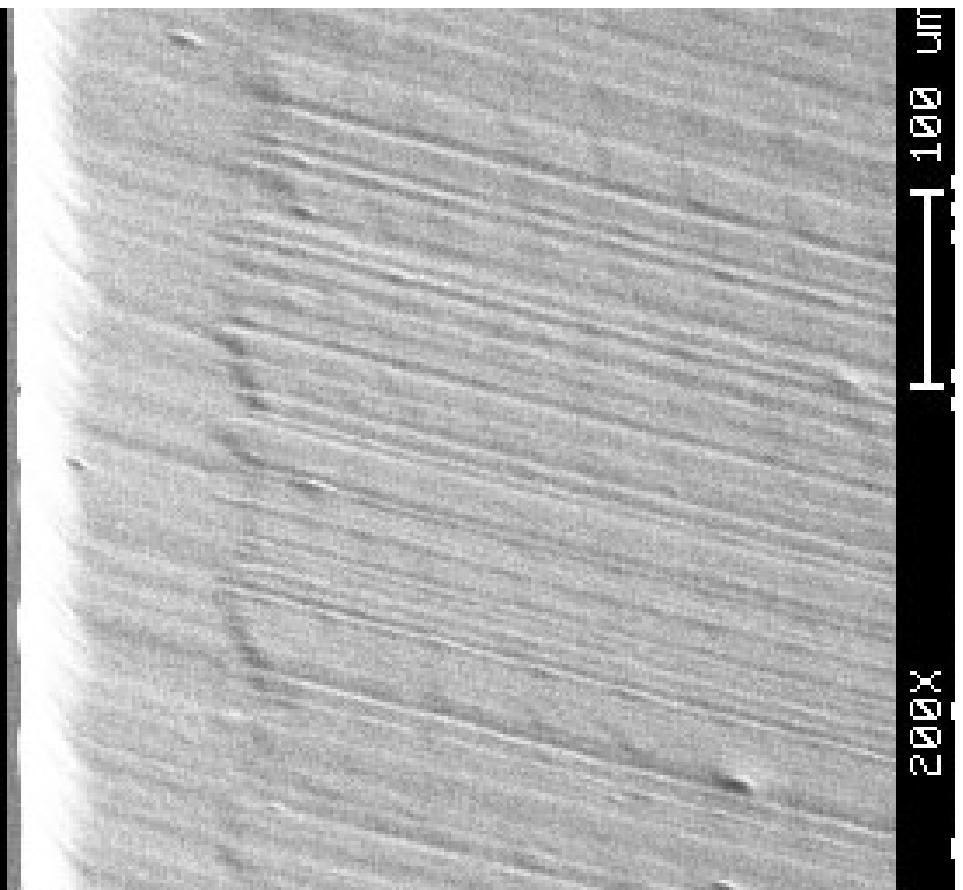}
\end{minipage}
\caption{Surface of the edge, cell side, of a horn after machining (left)
and after 60" etch and thermal processing}
\vspace{-0.2cm}
\label{BfrRFprocs}
\end{figure}

Following our autopsy many studies and simulation led to the
decision of rounding the edge to 3mm instead of 76~$\mu m$ radius
in order to avoid the thermal fatigue roughening mechanism. It is
believed that the new design of the IC and OC implementing this
rounding will cure our breakdown problem in those couplers. Also,
machining a 3mm edge radius should leave the surface extremely
smooth. We have shown on an identical input coupler of T53Vg3R,
Fig.\ref{BfrRFprocs}, that the 4 edges of the cavity side of the
horns present a roughness after machining. Following the etching
and thermal processing, that sees the structure, do not removed
this roughness. In a contrary, the edge on the waveguide side of
the horns are well finished.

In order to confirm the sharp corner problem two other structures
will be cut open. The T53Vg3RA has also reached the desired
73~MV/m but the IC were responsible of more than 75\% of the
breakdown Fig.\ref{T53Vg3}. 
It is then important to compare the two structures. Finally the
wakefield damping slots in a cell, of an NLC type RF-structure,
DDS3 cf Fig.\ref{historyOperation}, will be cut open and analyse.
Those damping slots might have sharp and not de-burred corners
leading to the understanding of DDS3 poor performance.


\begin{thebibliography}{1}

\bibitem{dolgashev:PAC01}
{V. Dolgashev, S. Tantawi}.
\newblock {Simulations of currents in X-band accelerator structures using 2D
  and 3D particle-in-cell code}.
\newblock In {\em {PAC 2001}}, 2001.

\bibitem{dolga:01}
{V. Dolgashev private communication}.

\bibitem{lepimpec:ISG8}
{F.~Le Pimpec, S.~Harvey, R.~Kirby, F.~Marcelja }.
\newblock {Results from the surface analysis of the autopsy of the T53VG3R
  structure}.
\newblock In {\em {ISG8}}, 2002.

\bibitem{mross:LC02}
{M. Ross et al.}
\newblock {Measurement of RF Breakdown in X-band Structures}.
\newblock In {\em {LC02 Workshop SLAC}}, 2002.

\bibitem{dolgashev:isg8}
{V. Dolgashev et al}.
\newblock {SW Processing, Pulse Heating}.
\newblock In {\em {ISG8}}, 2002.

\bibitem{pritzkau}
D.~Pritzkau.
\newblock {\em {thesis : RF Pulsed Heating}}.
\newblock SLAC-report-577.

\bibitem{berman:92}
A.~Berman.
\newblock {\em {Vacuum Engineering Calculations, Formulas, and Solved
  Exercises}}.
\newblock Academic Press, 1992.

\bibitem{sciffmann:93}
{K.I. Sciffmann}.
\newblock {\em Nanotechnology}, 4:163, 1993.


\bibitem{Singh:80}
{B. Singh et al}.
\newblock {AES study of sulfur surface segregation on polycrystalline copper}.
\newblock {\em JVST}, 17(1):29--33, 1980.

\end{thebibliography}

\end{document}